\documentclass{aa}  

\usepackage{graphicx}
\usepackage{txfonts}
\usepackage{xcolor}
\usepackage{hyperref}
\hypersetup{colorlinks, linkcolor=blue, citecolor=blue, urlcolor=blue}
\usepackage{amsmath}
\usepackage{ulem}
\usepackage{soul}
\usepackage{enumitem}

\newcommand{\pivec}{\mbox{\boldmath $\pi$}}

\newcommand{\te}{t_{\rm E}}
\newcommand{\thetae}{\theta_{\rm E}}

\newcommand{\pie}{\pi_{\rm E}}
\newcommand{\pien}{\pi_{{\rm E},N}}
\newcommand{\piee}{\pi_{{\rm E},E}}
\newcommand{\dl}{D_{\rm L}}
\newcommand{\ds}{D_{\rm S}}

\definecolor{brown}{rgb}{0.59, 0.29, 0.0}
\definecolor{darkgreen}{rgb}{0.0, 0.42, 0.24}
\definecolor{darkblue}{rgb}{0.01, 0.31, 0.59}
\definecolor{darkblue}{rgb}{0.0, 0.25, 0.42}
\definecolor{blue}{rgb}{0.0,0.0,1.0}
\definecolor{green}{rgb}{0.0,1.0,0.0}

\begin{document}

\title{KMT-2018-BLG-1743: Planetary Microlensing Event Occurring on Two Source Stars}

\author{
     Cheongho~Han\inst{1} 
\and Michael~D.~Albrow\inst{2}   
\and Sun-Ju~Chung\inst{3,4}      
\and Andrew~Gould\inst{5,6}      
\and Kyu-Ha~Hwang\inst{3} 
\and Youn~Kil~Jung\inst{3} 
\and Doeon~Kim\inst{1}
\and Hyoun-Woo~Kim\inst{3} 
\and Chung-Uk~Lee\inst{3} 
\and Yoon-Hyun~Ryu\inst{3} 
\and In-Gu~Shin\inst{3} 
\and Yossi~Shvartzvald\inst{7}  
\and Jennifer~C.~Yee\inst{8}    
\and Weicheng~Zang\inst{9}     
\and Sang-Mok~Cha\inst{3,10} 
\and Dong-Jin~Kim\inst{3} 
\and Seung-Lee~Kim\inst{3,4} 
\and Dong-Joo~Lee\inst{3} 
\and Yongseok~Lee\inst{3,10} 
\and Byeong-Gon~Park\inst{3,4} 
\and Richard~W.~Pogge\inst{6}
\\
(The KMTNet Collaboration)
}

\institute{
     Department of Physics, Chungbuk National University, Cheongju 28644, Republic of Korea  \\ \email{\color{blue} cheongho@astroph.chungbuk.ac.kr}     
\and University of Canterbury, Department of Physics and Astronomy, Private Bag 4800, Christchurch 8020, New Zealand                                     
\and Korea Astronomy and Space Science Institute, Daejon 34055, Republic of Korea                                                                        
\and Korea University of Science and Technology, 217 Gajeong-ro, Yuseong-gu, Daejeon, 34113, Republic of Korea                                           
\and Max Planck Institute for Astronomy, K\"onigstuhl 17, D-69117 Heidelberg, Germany                                                                    
\and Department of Astronomy, The Ohio State University, 140 W. 18th Ave., Columbus, OH 43210, USA                                                       
\and Department of Particle Physics and Astrophysics, Weizmann Institute of Science, Rehovot 76100, Israel                                               
\and Center for Astrophysics $|$ Harvard \& Smithsonian 60 Garden St., Cambridge, MA 02138, USA                                                          
\and Department of Astronomy, Tsinghua University, Beijing 100084, China                                            
\and School of Space Research, Kyung Hee University, Yongin, Kyeonggi 17104, Republic of Korea                                                           
}
\date{Received ; accepted}

\abstract
{}
{
We present the analysis of the microlensing event KMT-2018-BLG-1743.  The analysis was conducted 
as a part of the project, in which previous lensing events detected in and before the 2019 season 
by the KMTNet survey were reinvestigated with the aim of finding solutions of anomalous events 
with no suggested plausible models. 
}
{
The light curve of the event, with a peak magnification $A_{\rm peak}\sim 800$, exhibits two 
anomaly features, one around the peak and the other on the falling side of the light curve.  An 
interpretation with a binary lens and a single source (2L1S) cannot describe the anomalies.  By 
conducting additional modeling that includes an extra lens (3L1S) or an extra source (2L2S) 
relative to a 2L1S interpretation, we find that 2L2S interpretations with a planetary lens system 
and a binary source best explain the observed light curve with  $\Delta\chi^2\sim 188$ and $\sim 91$ 
over the 2L1S and 3L1S solutions, respectively.
Assuming that these $\Delta\chi^2$ values are adequate for distinguishing the models,
the event is the fourth 2L2S event and the second 2L2S planetary event.
The 2L2S interpretations are subject to a degeneracy, resulting in two solutions 
with $s>1.0$ (wide solution) and $s<1.0$ (close solution). 
}
{
The masses of the lens components and the distance to the lens are 
$(M_{\rm host}/M_\odot, M_{\rm planet}/M_{\rm J}, 
D_{\rm L}/{\rm kpc})
\sim (0.19^{+0.27}_{-0.111}, 0.25^{+0.34}_{-0.14}, 6.48^{+0.94}_{-1.03})$ 
and 
$\sim (0.42^{+0.34}_{-0.25}, 1.61^{+1.30}_{-0.97}, 6.04^{+0.93}_{-1.27})$ 
according to the wide and close solutions, respectively.  The source is a binary composed of an 
early G dwarf and a mid M dwarf.  The values of the relative lens-source proper motion expected 
from the two degenerate solutions, $\mu_{\rm wide}\sim 2.3$~mas~yr$^{-1}$ and  $\mu_{\rm close} 
\sim 4.1$~mas~yr$^{-1}$, are substantially different, and thus the degeneracy can be broken
by resolving the lens and source from future high-resolution imaging observations.
}
{}


\maketitle

\section{Introduction}\label{sec:one}

The standard light curve of a two-object (lens and source) lensing event with a single lens and 
a single source (1L1S) has a smooth and symmetric form \citep{Paczynski1986}.  Under the approximation 
of a rectilinear relative lens-source motion, the lensing light curve of a 1L1S event is described 
by three lensing parameters of $(t_0, u_0, \te)$ as
\begin{equation}
A = {u^2+2  \over u \sqrt{u^2+4}};\qquad
u= \left[ u_0^2 + \left( {t-t_0 \over \te} \right)^2\right]^{1/2},
\label{eq1}
\end{equation}
where the individual lensing parameters represent the time of the closest lens-source approach,
the lens-source separation (normalized to the angular Einstein radius $\thetae$) at $t_0$, and 
the event time scale, respectively.

For a fraction of lensing events, light curves exhibit deviations from the 1L1S form. Such 
deviations are most commonly caused by the duality of the lens \citep{Mao1991} or the source 
\citep{Griest1993}.  Hereafter, we refer to the three-object events with binary lens or binary 
source stars as 2L1S or 1L2S events, respectively.  At the time when such anomalous events were 
first found, for example, MACHO~LMC~1 \citep{Dominik1994} and OGLE~7 \citep{Udalski1994} 2L1S 
events and MACHO LMC 96-2 \citep{Becker1997} 1L2S event, interpreting the observed lensing light 
curves was a challenging task due to the difficulty of modeling caused by various technical issues.  
The first of these issues was the increased number of lensing parameters with the addition of the 
extra lens or source component, and this made it difficult to find a lensing solution, that is, 
a set of lensing parameters that best explain the observed light curves, via a grid approach. 
Second, the $\chi^2$ surface in the parameter plane was complex, especially for 2L1S events, due 
to the discontinuity in lensing magnifications caused by the formation of caustics, and this made 
it difficult to find a solution via a simple $\chi^2$ minimization approach.  Third, computing 
finite-source magnifications during the source star's crossing over the caustic required heavy 
computations, and this hampered prompt interpretations of anomalous lensing events.  The majority 
of these difficulties in the interpretation of multiple-object lensing events has been resolved 
with the adoption of sophisticated logic that is optimized for finding lensing solutions in a 
complex parameter space, such as the Markov Chain Monte Carlo (MCMC) method, together with the 
development of efficient methods for computing finite-source magnifications, such as the map-making 
method, for example, \citet{Dong2009}, the adaptive ray-shooting method, for example, \citet{Bennett2010}, 
and the image contour method, for example, \citet{Gould1997} and \citet{Bozza2018}.  As a result, 
three-object modelings are routinely being conducted for most anomalous events.

For a minor fraction of anomalous events, it is known that the observed light curves cannot be
explained by either a 2L1S or a 1L2S model. The difficulty of finding lensing solutions for 
these events suggests that interpreting their light curves requires more complex models than 
those including three objects. For some of these events, solutions including more than three 
objects were identified.  Currently, there exist 14 confirmed cases of lensing events with models 
including more than three objects.  See the list of these events in Table~1 of \citet{Han2021a}.  
However, there still exist events for which no plausible model has been proposed.

\begin{table}[t]
\small
\caption{Data and error rescaling factors\label{table:one}}
\begin{tabular*}{\columnwidth}{@{\extracolsep{\fill}}ccccc}
\hline\hline
\multicolumn{1}{c}{Data set}                           &
\multicolumn{1}{c}{$k$}                                &
\multicolumn{1}{c}{$\sigma_{\rm min}$ (mag)}                 &
\multicolumn{1}{c}{Time range (${\rm HJD^\prime}$)}    &
\multicolumn{1}{c}{$N_{\rm data}$}                     \\
\hline
KMTA   & 1.330   &  0.020    &  8170 -- 8400   &  193   \\
KMTC   & 1.175   &  0.020    &  8172 -- 8412   &  305   \\
KMTS   & 1.170   &  0.020    &  8177 -- 8401   &  192   \\
\hline
\end{tabular*}
\tablefoot{ ${\rm HJD}^\prime\equiv {\rm HJD}-2450000$.  
}
\end{table}

In this paper, we present the analysis of the planetary lensing event KMT-2018-BLG-1743. 
The event was reinvestigated in the project of reanalyzing anomalous events, for which lensing 
light curves could not be explained by either a 2L1S or a 1L2S model, among the previous 
lensing events detected in and before the 2019 season by the Korea Microlensing Telescope 
Network \citep[KMTNet:][]{Kim2016}.  This project has led to the discoveries of the 3L1S 
events OGLE-2018-BLG-1700 \citep{Han2020b} and OGLE-2019-BLG-0304 \citep{Han2021c}, in which 
the lenses have three components (planets in binary systems), the 2L2S event KMT-2019-BLG-0797 
\citep{Han2021a}, in which both the lens and source are binaries, and the 3L2S event KMT-2019-BLG-1715 
\citep{Han2021b}, in which the lens have three components and the source is composed of two stars.  
Here the notation ``3L'' indicates that the lens is a triple system.  The event KMT-2018-BLG-1743 
was known to be anomalous, but no detailed analysis has been presented due to the difficulty of 
explaining the anomalies with a three-object model.  In this work, we check the feasibility of 
interpreting the KMT-2018-BLG-1743 light curve with the introduction of an extra lens or source 
component.

For the presentation of the work, we organize the paper as follows.  In Sect.~\ref{sec:two}, 
we describe the observations of the lensing event and the data used in the analysis.  In 
Sect.~\ref{sec:three}, we model the lensing light curve under various interpretations and 
present the results of these models.  We specify the source type and estimate the angular 
Einstein radius of the event in Sect.~\ref{sec:four}.  We determine the physical parameters 
of the lens system in Sect.~\ref{sec:five}, and discuss the degeneracy in the interpretation 
of the event in Sect.~\ref{sec:six}.  We summarize the result of the analysis and conclude 
in Sect.~\ref{sec:seven}.

\section{Observations and data}\label{sec:two}

The source star of the lensing event KMT-2018-BLG-1743 is located toward the Galactic bulge
field. The equatorial and galactic coordinates of the source are 
$({\rm RA}, {\rm decl.})_{\rm J2000}=(18:17:08.01, -26:21:34.70)$ and 
$(l, b) = (5^\circ\hskip-2pt.694, -4^\circ\hskip-2pt.792)$, respectively. The baseline magnitude 
of the source before the lensing magnification was $I_{\rm base}=19.81$ according to the KMTNet 
scale.  We note that the source lies at a very dense field, toward which stellar images are, in 
most cases, heavily blended, and thus there exists, in general, little information about the stars 
of the field in public databases such as SIMBAD.\footnote{http://simbad.u-strasbg.fr/simbad/} The 
magnification of the source flux induced by lensing was found from the post-season investigation 
of the 2018 season KMTNet data using the AlertFinder System \citep{Kim2018}.

\begin{figure}
\includegraphics[width=\columnwidth]{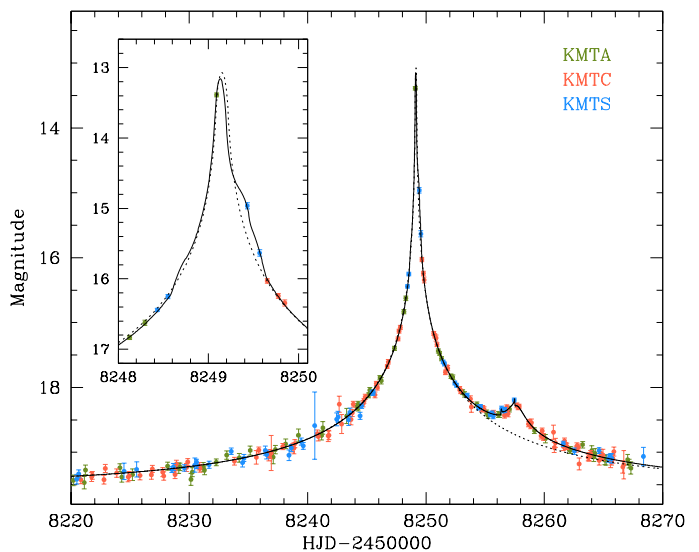}
\caption{
Light curve of the microlensing event KMT-2018-BLG-1743. The inset shows the zoom-in view of the 
peak region. The two curves drawn over the data points are the model curves of the 1L1S (dotted) 
and the wide 2L2S (solid) solutions. The colors of the telescopes in the legend are chosen to match 
those of the data points in the light curve.
}
\label{fig:one}
\end{figure}

The KMTNet group has conducted a microlensing survey since 2015 by utilizing three identical
KMTNet telescopes. These telescopes are globally distributed in three continents of the Southern
Hemisphere for the continuous coverage of lensing events: Australia (KMTA), Chile (KMTC), and
South Africa (KMTS).  Each telescope, with a 1.6~m aperture, is equipped with a camera providing 
a $2^\circ \times 2^\circ$ field of view.  The lensing event was located in the KMTNet BLG31 field, 
toward which observations were conducted with a 2.5~hr cadence.  Most images were acquired in the 
$I$ band, and about one tenth of images were obtained in the $V$ band.  The $I$-band data were 
used for the light curve analysis, and the $V$-band data were used for the measurement of the 
source color.

Reduction of the data was conducted using the pySIS code developed by \citet{Albrow2009}. 
The photometry code is a customized version of the difference imaging method \citep{Tomaney1996, Alard1998}, 
that was developed for the optimal photometry of stars located in a very dense star field.  
Additional photometry using the pyDIA software \citep{Albrow2017} was done for a subset of KMTC 
$I$- and $V$-band data to measure the color of the source and to construct the color-magnitude 
diagram (CMD) of stars around the source.  We will describe the detailed procedure of the source 
color measurement in Sect.~\ref{sec:four}.  Following the routine of \citet{Yee2012}, the error 
bars of the data used in the analysis were readjusted by $\sigma=[\sigma_{\rm min}^2+
(k\sigma_0)^2]^{1/2}$, where $\sigma_0$ denotes the error bar estimated by the automated pipeline, 
and $(\sigma_{\rm min}, k)$ represent the scatter of data and the rescaling factor used to make 
$\chi^2$ per degree of freedom become unity, respectively. In Table~\ref{table:one}, we present 
the data rescaling coefficients along with the time range and the number of data points, 
$N_{\rm data}$, for the individual data sets.

\begin{figure}
\includegraphics[width=\columnwidth]{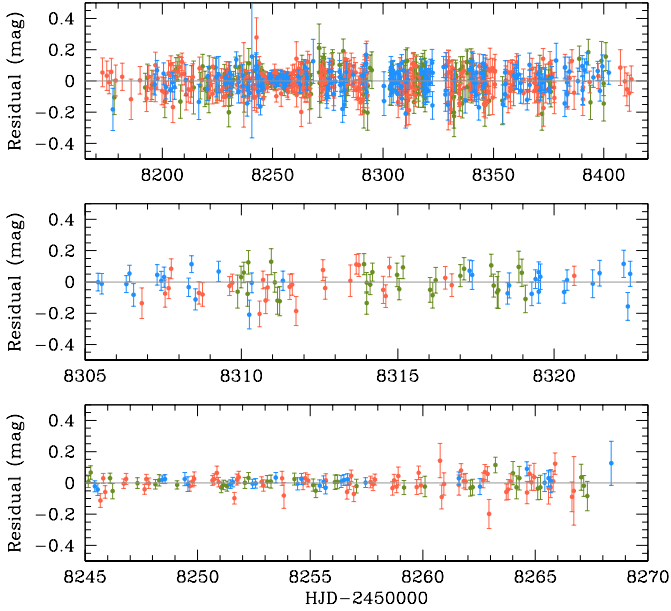}
\caption{
Residual from the best-fit model (wide 2L2S model) in various time ranges.  The top and 
middle panels show the residuals in the time ranges of $8165\leq {\rm HJD}^\prime \leq 8420$ 
(225~days) and $8305\leq {\rm HJD}^\prime \leq 8323$ (18~days), respectively.  The bottom 
panel shows the residual in the time range of $8245\leq {\rm HJD}^\prime \leq 8270$ (25~day), 
during which the source was substantially magnified.
}
\label{fig:two}
\end{figure}

Figure~\ref{fig:one} shows the lensing light curve of KMT-2018-BLG-1743.  Inspection of the 
light curve reveals three characteristics. First, the light curve exhibits an anomaly in the 
falling side of the light curve. The anomaly is centered at ${\rm HJD}^\prime\equiv {\rm HJD}
-2450000\sim 8258$ (2018-05-19), and lasted for about 6 days.  Second, the event was highly 
magnified. A 1L1S modeling conducted with the exclusion of the data around the anomaly in the 
wing yields $(t_0, u_0, \te, \rho)\sim (8249.150, 5\times 10^{-6}, 28.18~{\rm days}, 2.46
\times 10^{-3})$.  Here $\rho$ denotes the normalized source radius, which is defined by the 
ratio of the angular source radius, $\theta_*$, to the angular Einstein radius, $\thetae$, 
that is, $\rho=\theta_*/\thetae$, and it is included in the modeling to account for possible 
finite-source effects that may occur when the lens passes over the surface of the source. 
In computing finite magnifications, we consider limb-darkening effects by adopting a linear 
coefficient of $u=0.5$ from \citet{Claret2000}, considering that the source is an early G-type 
dwarf.  The detailed procedure of the source type specification is described in Sect.~\ref{sec:four}.
The 1L1S model curve (dotted curve) is 
drawn over the data points in Figure~\ref{fig:one}. The magnification at the peak of the light 
curve estimated from the 1L1S model is $A_{\rm peak}\sim 800$.  Third, the light curve exhibits 
an additional anomaly in the peak region as well as the anomaly in the wing.  To better show 
this central deviation, we present the zoom-in view of the peak region in the inset of
Figure~\ref{fig:one}. The central deviation is most evident for the two KMTS data points taken 
at the epochs of ${\rm HJD}^\prime=8249.430$ and 8249.568, which exhibit deviations of 
$\Delta I=0.45$~mag and 0.21~mag from the 1L1S model, respectively. These deviations are much 
greater than the photometric uncertainties of nearby data points.

We checked the possibility of systematics in the data and the variability of the source or blend
by inspecting the residual from the best-fit model that will be described in the following section.
The residual is shown in Figure~\ref{fig:two}.  The top panel, with a time range of $8165\leq 
{\rm HJD}^\prime \leq 8420$ (225~days), and the middle panel, with $8305\leq {\rm HJD}^\prime 
\leq 8323$ (18~days), are presented to check long- and short-term systematics or variability in 
the data.  The bottom panel, showing the residual in the time range of $8245\leq {\rm HJD}^\prime 
\leq 8270$ (25~day), during which the source was substantially magnified, is shown to check the 
possible systematics that might arise in the measurement of  the light variation.  It is found that 
the data in all inspected time ranges do not show any symptom of systematics or source variation.

\begin{figure}[t]
\includegraphics[width=\columnwidth]{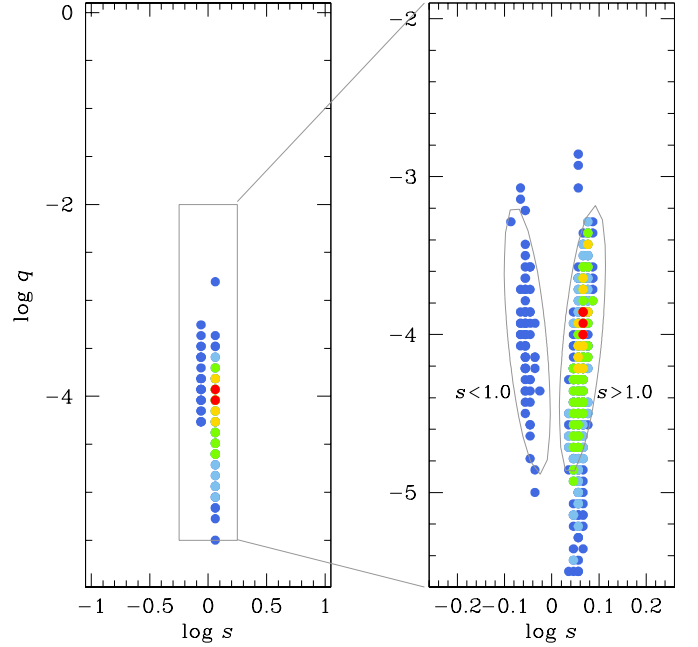}
\caption{
$\Delta\chi^2$ map in the $\log s$--$\log q$ parameter plane obtained from the 2L1S modeling. 
Points marked in different colors represent the regions with 
$\Delta\chi^2\leq n(1^2)$ (red), 
$\Delta\chi^2\leq n(2^2)$ (yellow), 
$\Delta\chi^2\leq n(3^2)$ (green), 
$\Delta\chi^2\leq n(4^2)$ (cyan), and 
$\Delta\chi^2\leq n(5^2)$ (blue), where $n=5$. 
The left panel shows the map in the region with $-1.0<\log s\leq 1.0$ and 
$-5.8<\log q\leq 0.0$, and the right panel shows the map in the narrower region with 
$-0.26<\log s\leq 0.26$ and $-5.6<\log q\leq -1.9$.
}
\label{fig:three}
\end{figure}

An event with a very high magnification is an important target for follow-up observations due to
the high chance of detecting planet-induced perturbations in the peak region of the light curve 
\citep{Griest1998}. Despite the very high magnification, no followup observation was conducted 
for KMT-2018-BLG-1743, because the KMTNet real-time AlertFinder \citep{Kim2018} only began 
operation on 21 June 2018, that is, 40 days after the peak. Moreover, the AlertFinder was not 
applied to field BLG31 until 2019.

\section{Interpretation of the anomaly}\label{sec:three}

A key for the interpretation of the event is to explain both anomalies, that is, the one in the 
peak and the other in the wing of the observed light curve.  Hereafter, we refer to the individual 
anomalies as the central and peripheral anomalies, respectively.  For the interpretation of the 
anomalies, we test various lensing models under 2L1S, 3L1S, and 2L2S interpretations.  The 2L1S 
model is tested because a binary lens with a very low-mass companion can generate two short-term 
anomaly features in the lensing light curve under a certain lens system configuration.  As already 
mentioned and to be fully discussed in Sect.~\ref{sec:three-one}, it is difficult to precisely 
describe the observed data with a 2L1S model.  The 3L1S and 2L2S models are tested to check whether 
the anomaly features can be described by introducing an additional lens or source component.  In 
the following subsections, we describe the procedures of the individual modelings and present the 
results of the analyses.

\begin{table}
\small
\caption{Lensing parameters of the 2L1S and 3L1S models\label{table:two}}
\begin{tabular*}{\columnwidth}{@{\extracolsep{\fill}}lll}
\hline\hline
\multicolumn{1}{c}{Parameter }          &
\multicolumn{1}{c}{2L1S model }         &
\multicolumn{1}{c}{3L1S model }         \\
\hline
$\chi^2$                      &   841.2                 &   743.8                     \\
$t_0$ (${\rm HJD}^\prime$)    &  $8249.139 \pm 0.003 $  &  $8248.631 \pm 0.052   $    \\
$u_0$ ($10^{-3}$)             &  $0.71 \pm 0.20      $  &  $32.45 \pm 3.68       $    \\
$\te$ (days)                  &  $31.26 \pm 1.35     $  &  $31.53 \pm 1.05       $    \\
$s_2$                         &  $1.155 \pm 0.007    $  &  $1.177 \pm 0.006      $    \\
$q_2$ ($10^{-3}$)             &  $0.09 \pm 0.02      $  &  $0.17 \pm 0.02        $    \\
$\alpha$ (rad)                &  $3.149 \pm 0.001    $  &  $3.149 \pm 0.003      $    \\
$s_3$                         &   --                    &  $0.118 \pm 0.007      $    \\
$q_3$                         &   --                    &  $0.515 \pm 0.101      $    \\
$\psi$ (rad)                  &   --                    &  $1.135 \pm 0.029      $    \\
$\rho$ ($10^{-3}$)            &  $1.80 \pm 0.19      $  &  $0.86 \pm 0.05        $    \\
\hline
\end{tabular*}
\tablefoot{ ${\rm HJD}^\prime\equiv {\rm HJD}-2450000$.  
}
\end{table}

\subsection{2L1S interpretation}\label{sec:three-one}

In principle, a 2L1S lensing light curve can produce two anomaly features.  This is possible 
because a binary lens with a very small mass ratio between the lens components, $M_1$ and 
$M_2< M_1$, such as a binary pair composed of a planet and a host, induces two sets of caustics, 
in which one is located close to $M_1$ (central caustic) and the other is located away from $M_1$ 
(planetary caustic).  Then, two anomaly features can arise when a source passes both the central 
and planetary caustics.

Keeping this possibility in mind, we conducted a 2L1S modeling of the light curve. In addition to
the 1L1S parameters, a 2L1S modeling requires one to include additional parameters to describe
the lens binarity. These parameters are $(s, q, \alpha)$, which denote the projected $M_1$--$M_2$ 
separation (normalized to $\thetae$), the mass ratio $q=M_1/M_2$, and the angle between the source 
trajectory and the $M_1$--$M_2$ axis (source trajectory angle). The modeling was carried out in two 
steps. In the first step, we conducted grid searches for $s$ and $q$ with multiple starting points 
of $\alpha$ evenly distributed in the range of $0< \alpha\leq 2\pi$.  In this procedure, the lensing 
parameters $(t_0, u_0, \te)$ were searched for using a downhill approach based on the MCMC method. 
Besides these parameters, a lensing modeling requires one to include two flux parameters $(f_{s,i}, 
f_{b,i})$ for each observatory, and these parameters are obtained from the regression of the observed 
flux, $F_i$, to the model by $F_i(t) = f_{s,i}A(t)+f_{b,i}$.  We constructed a $\Delta\chi^2$ map in 
the $s$--$q$ plane from the modeling, and identified local solutions in the $\Delta\chi^2$ map.  
Figure~\ref{fig:three} shows the $\Delta\chi^2$ map in the $\log s$--$\log q$ plane obtained in this 
procedure.  In the second step, we conducted an additional modeling to refine the individual local 
solutions found in the first step by allowing all parameters to vary. This two-step procedure allows 
us to identify degenerate solutions, it they exist.

\begin{figure}[t]
\includegraphics[width=\columnwidth]{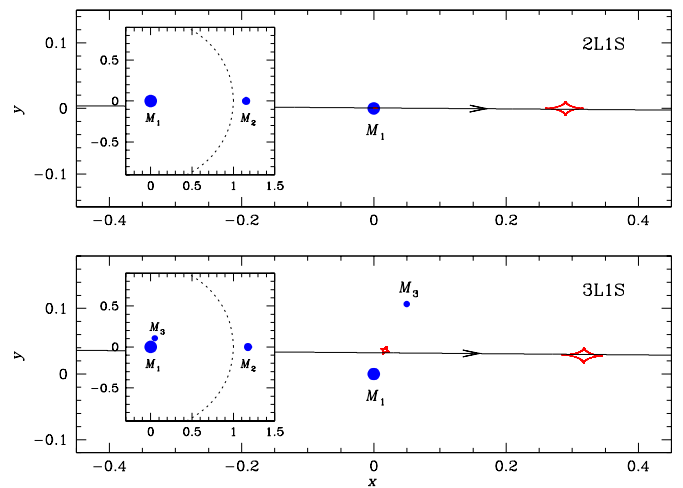}
\caption{
Lens system configurations of the 2L1S (upper panel) and 3L1S (lower panel) models. In each 
panel, the inset shows the broad region including the positions of the lens components: blue 
dots marked by $(M_1, M_2)$ for the 2L1S model and $(M_1, M_2, M_3)$ for the 3L1S model.  The 
dotted circle in each inset represents the Einstein ring.  The source motion is represented by 
a line with an arrow. The red closed figures represent the caustics.
}
\label{fig:four}
\end{figure}

The modeling yielded a predicted solution: a source passing both the central and planetary caustics 
produced by a planetary lens system.  The lensing parameters of the best-fit 2L1S model are listed 
in Table~\ref{table:two} along with the $\chi^2$ value of the fit. We note that the binary parameters 
are represented as $(s_2, q_2)$ to distinguish them from the parameters of a possible third lens mass, 
$(s_3, q_3)$, to be discussed in the following subsection. The estimated binary parameters are 
$(s_2, q_2)\sim (1.16, 9\times 10^{-5}$), which would indicate that the event was produced by a 
planetary system containing a planet with a very low planet-to-host mass ratio and a projected 
separation slightly larger than the Einstein ring of the host. The lens system configuration, 
showing the source trajectory with respect to the positions of the lens components and the resulting 
caustics, is shown in the upper panel of Figure~\ref{fig:four}. The configuration shows that the 
central caustic is located very close to the host ($M_1$), and the planetary caustic is located at 
a position with a separation $s-1/s\sim 0.3$ from $M_1$ on the planet ($M_2$) side with respect to 
the host.  As expected, the source passes both the central and planetary caustics, and this produces 
the central and peripheral anomalies, respectively. We note that two anomaly features can also be 
produced by a close planet with a separation less than $\thetae$, that is, $s<1.0$.  This local 
solution appears in the $\Delta\chi^2$ map presented in Figure~\ref{fig:three}.  In this case, the 
induced planetary caustics have a different shape and number from those of the caustic induced by a 
wide planet with $s>1.0$ \citep{Han2006}. We found that the solution with a source trajectory connecting 
the planetary caustic induced by a planet with $s<1.0$ and the central caustic results in a substantially 
worse fit, by $\Delta\chi^2=340.5$, than the presented solution with $s>1.0$.

\begin{figure}
\includegraphics[width=\columnwidth]{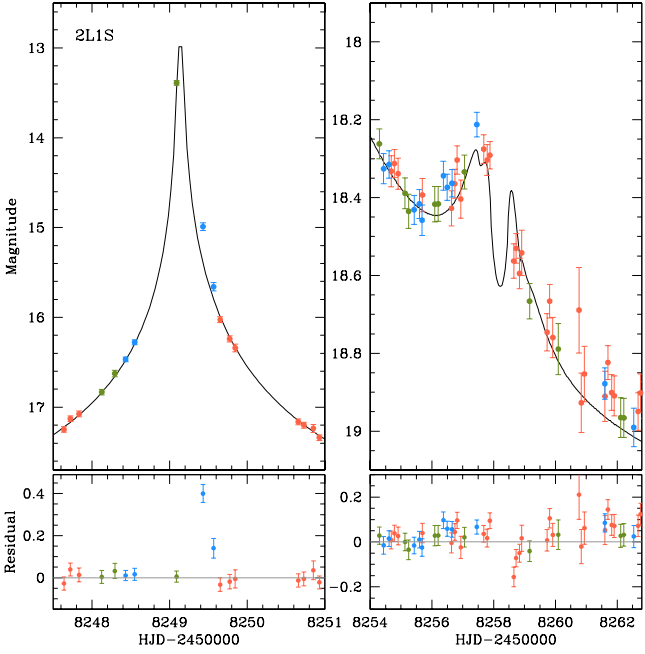}
\caption{
Model curve of the 2L1S solution and the residuals from the model in the central (left panels) and 
peripheral (right panels) anomaly regions.
}
\label{fig:five}
\end{figure}

It is found that the 2L1S model is not adequate for the precise description of the data, and its 
fit to the data is worse than the fit of the model to be discussed in Sect.~\ref{sec:three-three} 
by $\Delta\chi^2\sim 188$.  To demonstrate this inadequacy, in the left and right panels of 
Figure~\ref{fig:five}, we present the 2L1S model curve and the residuals from the model in the 
central and peripheral anomaly regions, respectively.  From the comparison of the data and the 
model in the central anomaly region, it is found that the two KMTS data points at ${\rm HJD}^\prime 
=8249.430$ and 8249.568 still exhibit considerable deviations from the model. It is also found that 
an important fraction of data points in the peripheral anomaly region lie outside the error bars 
from the model. The inadequacy of the model in describing the observed data suggests that a different 
interpretation of the event is needed.

We additionally check whether the deviations from the static 2L1S model can be explained by 
higher-order effects, especially the orbital motion of the lens.  We check the higher-order 
effects because there have been two cases, in which an extra lens body was incorrectly inferred 
due to the omission of the lens orbital motion in the 2L1S modeling together with the sparse 
coverage of the grid parameter ($s$, $q$, $\alpha$) space: MACHO-97-BLG-41 \citep{Bennett1999, 
Albrow2000, Jung2013} and OGLE-2013-BLG-0723 \citep{Udalski2015, Han2016}.  This check was done 
in two steps.  In the first step, we explore the grid parameter space with an increased density 
by doubling the resolution of the grid in each dimension.  It is found that this does not yield 
any local solution other than the solution presented in Table~\ref{table:two}.  In the second step, 
we consider the lens-orbital and microlens-parallax effects so that the source trajectory can be 
non-rectilinear.  Considering these higher-order effects requires one to include four additional 
parameters: $\pien$, $\piee$, $ds/dt$, and $d\alpha/dt$.  The first two parameters represent the 
north and east components of the microlens-parallax vector $\pivec_{\rm E}$, respectively, and the 
other two parameters indicate the change rates of the binary separation and the source trajectory 
angle, respectively.  The model with the higher-order effects improves the fit by $\Delta\chi^2=51.4$ 
with respect to the static model, but the fit is worse than the best model (in Sect.~\ref{sec:three-three}) 
by $\Delta\chi^2=136.8$, still leaving significant deviations in the central anomaly region.  Furthermore, 
the resulting value of the microlens parallax $\pie\sim 1.25$ is absurdly high, yielding a primary lens 
mass of $M_1\sim 0.02~M_\odot$, which is very unusual considering the very usual event time scale of 
$\te\sim 31$~days.  All these facts indicate that the residual from the 2L1S model is not ascribed 
to high-order effects, and the unusual higher-order parameters are induced by the incorrect basis 
static model.

\begin{figure}
\includegraphics[width=\columnwidth]{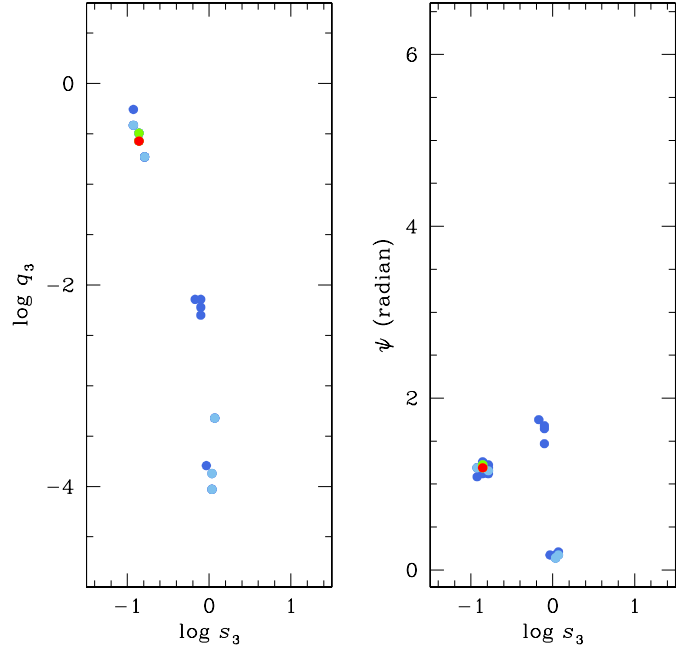}
\caption{
$\Delta\chi^2$ maps in the $\log s_3$--$\log q_3$ (left panel) and $log s_3$--$\psi$ (right panel) 
obtained from the 3L1S grid search. The color coding is same as that in Fig.~\ref{fig:three}, 
except that $n=1$.
}
\label{fig:six}
\end{figure}

\subsection{3L1S interpretation}\label{sec:three-two}

Not being able to find a 2L1S solution that precisely describes the observed light curve, we 
additionally examined a 3L1S model.  We tested this model because the major deviation from the 
2L1S model appeared in the central magnification region. If the lens has a third mass, $M_3$, 
the central caustic induced by $M_3$ may affect the magnification pattern of the central 
region \citep{Gaudi1998, Han2005}, and this may explain the deviation from the 2L1S model in the 
central anomaly. A modeling considering a third lens component requires one to include additional 
parameters of $(s_3, q_3, \psi)$, which represent the separation and mass ratio between $M_1$ and 
$M_3$, that is, $q_3=M_3/M_1$, and the orientation angle of $M_3$ as measured from the $M_1$--$M_2$ 
axis with its origin at $M_1$, respectively.

The 3L1S modeling was carried out in four steps.
\begin{enumerate}[label={(\arabic*)}]
\item
In the first step, we conducted a 2L1S modeling of the light curve with the exclusion of the 
data in each of the central and peripheral anomalies. This yielded three sets of 2L1S solutions, 
in which one was obtained with the exclusion of the data around the central anomaly 
($8248.0\leq {\rm HJD}^\prime \leq 8250.0$), and the other two solutions were obtained with the 
exclusion of the data around the peripheral anomaly ($8252.0\leq {\rm HJD}^\prime \leq 8270.0$).
\item
In the second step, we conducted a 3L1S modeling, in which the parameters $(s_3, q_3, \psi)$ were
searched for using a grid approach, and the other parameters were searched for using a downhill 
approach. In this step, we fix the parameters related to $M_2$, that is, $(s_2, q_2, \alpha)$, as 
the values of the solutions found from the 2L1S modeling conducted in the first step.
\item
In the final step, we refined the solutions found from the second step by releasing all parameters
as free parameters. 
\item
We repeated the procedure (2) and (3) for the three sets of the 2L1S solutions found in the step (1).
\end{enumerate}
It was found that the 3L1S solution starting from the $M_2$ parameters obtained with the exclusion 
of the data around the central anomaly yielded the best-fit solution.  
Figure~\ref{fig:six} shows the $\Delta\chi^2$ maps in the 
$\log s_3$--$\log q_3$ and $\log s_3$--$\psi$ planes obtained from the grid modeling step (2).
The solutions resulting from the 2L1S solutions with the exclusion of the peripheral anomalies were 
worse than the best-fit model by $\Delta\chi^2 \gtrsim 90$.

\begin{figure}
\includegraphics[width=\columnwidth]{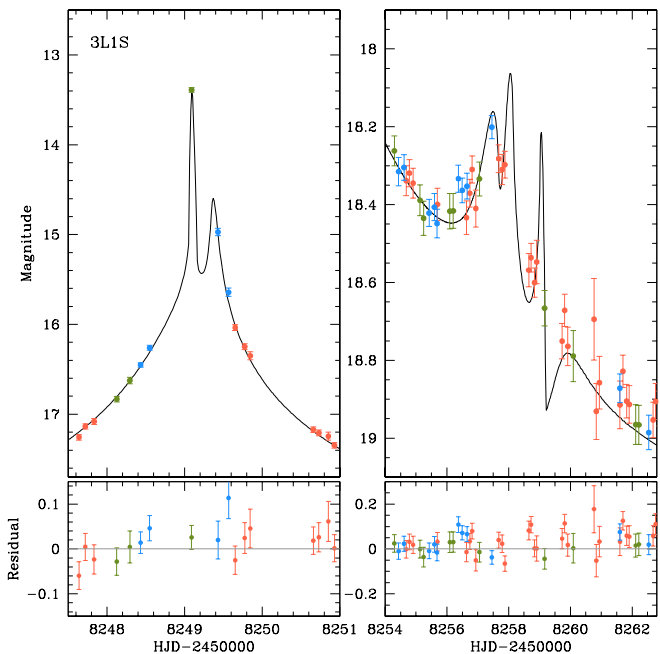}
\caption{
Model curve of the 3L1S solution and the residuals from the model. Notations are same as 
those in Fig.~\ref{fig:five}.
}
\label{fig:seven}
\end{figure}

Figure~\ref{fig:seven} shows the model curve and residuals from the best-fit 3L1S model in the 
central and peripheral anomaly regions. The lensing parameters of the solution are listed in 
Table~\ref{table:two} together with the $\chi^2$ value of the fit. It is found that the parameters 
of the 3L1S model related to the $M_1$--$M_2$ pair are similar to those of the 2L1S model, and the 
parameters related to the additional lens component $M_3$ are $(s_3, q_3, \psi)\sim (0.12, 0.51, 
65^\circ)$.  In the lower panel of Figure~\ref{fig:four}, we present the lens system configuration 
for the 3L1S solution.  The configuration is similar to that of the 2L1S solution, except that 
the extra lens component $M_3$ induces a tiny astroid-shape caustic around $M_1$. The source passes 
through the central caustic induced by $M_3$, and this substantially reduces the 2L1S residuals, 
especially in the central anomaly region. The comparison of the fit with that of the 2L1S model, 
presented in Figure~\ref{fig:five}, indicates that the 3L1S model provides a better fit than the 
2L1S solution, by $\Delta \chi^2=97.4$.  Although the fit of the 3L1S model looks fairly good, we 
reserve judgment the model until we test an additional interpretation in the following subsection.

\begin{figure}
\includegraphics[width=\columnwidth]{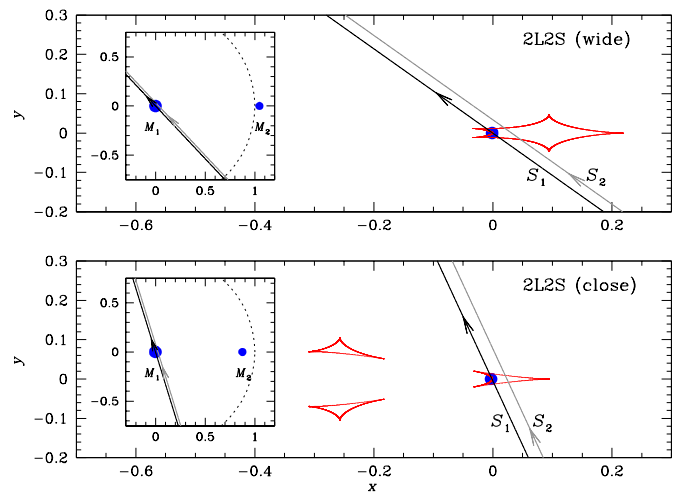}
\caption{
Lens system configurations of the wide (upper panel) and close (lower panel) 2L2S models.  
Notations are same as those in Fig.~\ref{fig:four}, except that there are two source trajectories 
for the primary ($S_1$) and the secondary ($S_2$) source stars.  The offset between the arrows 
on the two source-star trajectories indicates the relative position of the two sources.
}
\label{fig:eight}
\end{figure}

\begin{table}
\small
\caption{Lensing parameters of the 2L2S model\label{table:three}}
\begin{tabular*}{\columnwidth}{@{\extracolsep{\fill}}lll}
\hline\hline
\multicolumn{1}{c}{Parameter }          &
\multicolumn{1}{c}{Wide }         &
\multicolumn{1}{c}{Close }         \\
\hline
$\chi^2$                      &   653.0                  &    657.8                    \\
$t_{0,1}$ (HJD$^\prime$)      &  $8249.126\pm 0.006   $  &   $8249.119 \pm 0.013  $    \\
$u_{0,1}$ ($10^{-3}$)         &  $0.25 \pm 0.51       $  &   $1.51 \pm 0.35       $    \\
$t_{0,2}$ (HJD$^\prime$)      &  $8257.596 \pm 0.054  $  &   $8257.579 \pm 0.048  $    \\
$u_{0,2}$                     &  $-0.023 \pm 0.003    $  &   $-0.022 \pm 0.003    $    \\
$\te$ (days)                  &  $28.03 \pm 1.21      $  &   $27.76 \pm 1.21      $    \\
$s$                           &  $1.048 \pm 0.006     $  &   $0.878 \pm 0.016     $    \\
$q$ ($10^{-3}$)               &  $1.21 \pm 0.21       $  &   $3.68 \pm 0.65       $    \\
$\alpha$ (rad)                &  $0.820 \pm 0.068     $  &   $1.274 \pm 0.075     $    \\
$\rho_1$ ($10^{-3}$)          &  $2.08 \pm 0.38       $  &   $1.18 \pm 0.45       $    \\
$\rho_2$ ($10^{-3}$)          &   --                     &    --                       \\
$q_{F,I}$                     &  $0.066 \pm 0.005     $  &   $0.073 \pm 0.005     $    \\
\hline
\end{tabular*}
\tablefoot{ ${\rm HJD}^\prime\equiv {\rm HJD}-2450000$.  
}
\end{table}

\subsection{2L2S interpretation}\label{sec:three-three}

We additionally checked the possibility that both the lens and source were binaries.  We examined 
this model because 2L2S and 3L1S models occasionally yield similar light curves, as demonstrated 
in the cases of the lensing events KMT-2019-BLG-1953 \citep{Han2020a} and KMT-2019-BLG-0797 
\citep{Han2021a}.  A modeling with the inclusion of an extra source star, $S_2$, requires one to 
include additional parameters. These parameters are $(t_{0,2}, u_{0,2}, \rho_2, q_F)$, which 
represent the time and separation of $S_2$ at the closest approach to a reference position of the 
lens, the normalized radius of $S_2$, and the flux ratio between the source stars, respectively. 
We use the notations of the lensing parameters  related to the first source, $S_1$, as $(t_{0,1}, 
u_{0,1}, \rho_1)$ to distinguish them from the parameters related to the second source.  We started 
the 2L2S modeling with the three solutions obtained from the 2L1S modeling with the exclusion of the 
data around each of the central and peripheral anomalies.  We then set the initial values of the 
parameters $(t_{0,2}, u_{0,2}, \rho_2, q_F)$ considering the time and magnitude of the anomaly that 
was excluded in the initial 2L1S modeling.

The 2L2S modeling yielded two solutions that well described the data in both the central and
peripheral anomalies. In Table~\ref{table:three}, we list the lensing parameters and $\chi^2$ 
values of the fits for the two 2L2S solutions. It is found that one solution has a binary lens 
separation greater than unity ($s\sim 1.05$), and the other solution has a separation less than 
unity ($s\sim 0.88$). We refer to the solutions with $s>1.0$ and $s<1.0$ as ``wide'' and ``close'' 
solutions, respectively. The estimated mass ratio between the lens components is $q_{\rm wide}\sim 
1.2\times 10^{-3}$ for the wide solution and $q_{\rm close}\sim 3.7\times 10^{-3}$ for the close 
solution, and thus the mass of the companion is in the planetary-mass regime regardless of the 
solutions. We note that the binary separations of the pair of the degenerate solutions are not 
in the relation of $s_{\rm close}\sim 1/s_{\rm wide}$, and the source trajectory angles of the 
two solutions, $\alpha_{\rm wide}\sim 47^\circ$ and $\alpha_{\rm close}\sim 73^\circ$, are 
substantially different from each other.  This indicates that the degeneracy between the two 
solutions is not caused by the well-known close--wide degeneracy arising due to the similarity 
between the central caustics of the close and wide binaries with $s$ and $s^{-1}$ \citep{Griest1998, 
Dominik1999}.  Rather, it is  caused by the lack of observations during the time interval 
$8248.6 \lesssim {\rm HJD}^\prime \lesssim 8249.6$, when the wide and close solutions have 
single-peak and double-peak morphologies, respectively.  For both solutions, the secondary 
source is fainter than the primary source with an $I$-band flux ratio of $q_F\sim 7\%$.

\begin{figure}[t]
\includegraphics[width=\columnwidth]{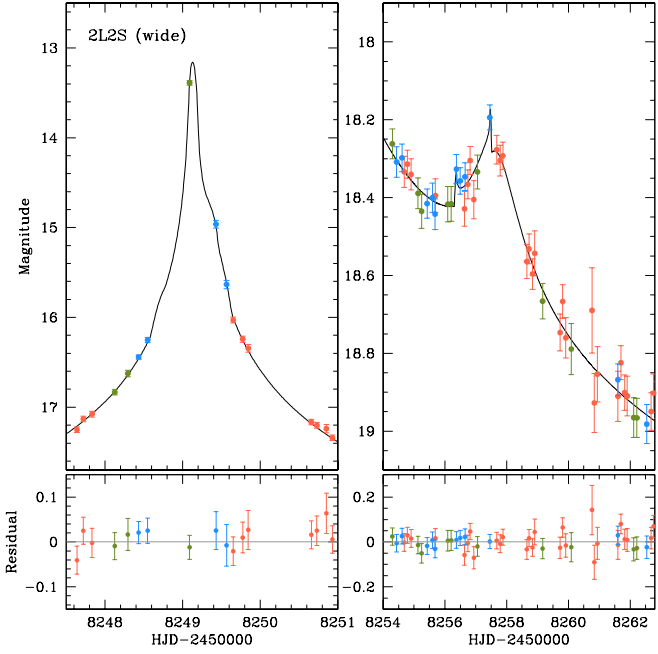}
\caption{
Model curve of the wide 2L2S solution and the residuals from the model.  Notations are same as 
those in Fig.~\ref{fig:five}.
}
\label{fig:nine}
\end{figure}

\begin{figure}[t]
\includegraphics[width=\columnwidth]{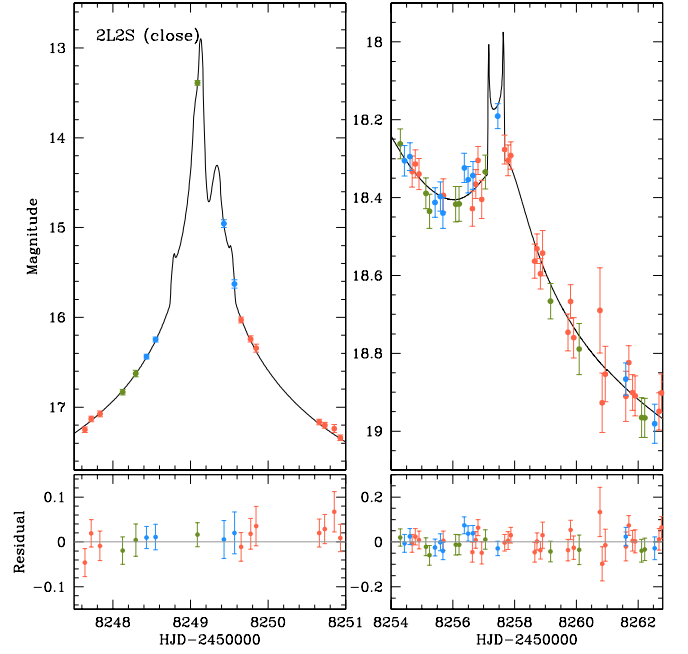}
\caption{
Model curve of the close 2L2S solution and the residuals from the model.  Notations are same 
as those in Fig.~\ref{fig:four}.
}
\label{fig:ten}
\end{figure}

Figure~\ref{fig:eight} shows the lens system configurations of the two 2L2S solutions. The 
trajectories of the primary and secondary source stars are marked by $S_1$ and $S_2$, respectively. 
In each configuration, the tips of the arrows on the source trajectories represent the positions of 
$S_1$ and $S_2$ at a same epoch, and thus the configuration indicates that $S_2$ trails $S_1$ for 
both the wide and close solutions.  Regardless of the solution, both the primary and secondary 
source stars cross the caustic, and the caustic crossings of $S_1$ and $S_2$ explain the central 
and peripheral anomalies, respectively. The model curves and the residuals from the models of the 
wide and close solutions in the regions of the anomalies are shown in Figures~\ref{fig:nine} and 
\ref{fig:ten}, respectively. Despite the significant difference in the lens system configurations, 
it is found that both solutions result in similar fits to the data.

\begin{figure}
\includegraphics[width=\columnwidth]{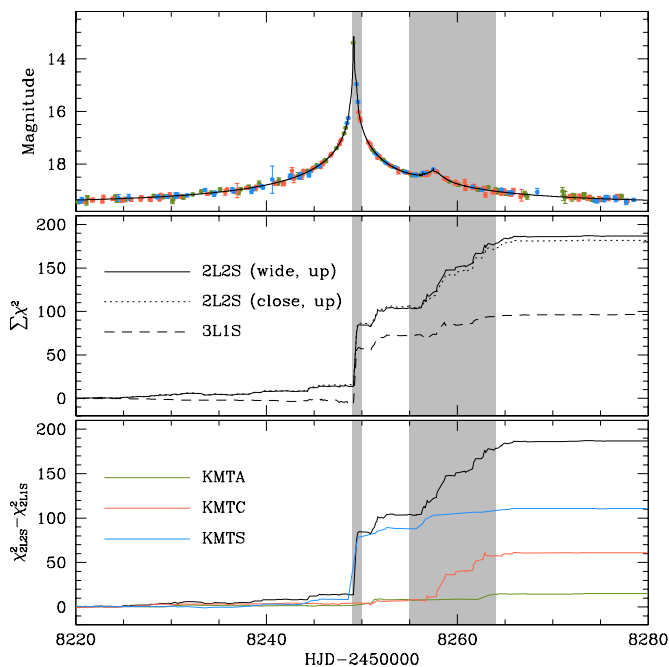}
\caption{
Cumulative distributions of $\Delta\chi^2$ from the 2L1S model for the 3L1S and 2L2S (wide and
close) models. The shaded area indicate the central (left) and peripheral (right) anomaly regions.
}
\label{fig:eleven}
\end{figure}

We find that the models obtained under the 2L2S interpretation provide substantially better fits 
to the observed data than the other models based on the 2L1S and 3L1S interpretations.  This can 
be seen in the middle panel of Figure~\ref{fig:eleven}, where we plot the cumulative distributions 
of the $\chi^2$ difference relative to the 2L1S model, that is, $\Delta\chi^2=\chi^2_{\rm 2L1S}-\chi^2$, 
for the 3L1S and 2L2S (wide and close) models.  We find that the fit of the wide (close) 2L2S model 
is better than the 3L1S and 2L1S models by $\Delta\chi^2=90.8$ (86.0) and 188.2 (184.3), respectively. 
As expected, the fit improvement occurs mainly in the anomaly regions.  The comparison of the fits 
between the two 2L2S solutions indicates that the wide solution slightly better describes the 
observed data in the peripheral anomaly region than does the close solution, but the $\chi^2$ 
difference is small, $\Delta\chi^2=4.8$, and thus we consider the close model as a viable solution.  
The bottom panel show the contribution to the 2L2S fit improvement, that is, $\Delta\chi^2_{\rm 2L2S}
=\chi^2_{\rm 2L1S}-\chi^2_{\rm 2L2S}$, by the individual data sets.  It shows that the contribution 
to $\Delta\chi^2_{\rm 2L2S}$ in the region around the central anomaly comes mostly from the KMTS 
data set, because the region is mainly covered by this data set, and the contribution in the 
peripheral anomaly region comes from both the KMTC and KMTS data sets.  The fit improvement by 
both data sets further supports the validity of the model.  On the other hand, the contribution to 
$\Delta\chi^2_{\rm 2L2S}$ by the KMTA data set is small due to its spare coverage of both anomaly 
regions.

With the superiority of the fit to the data over the other models, we conclude that KMT-2018-BLG-1743 
is a planetary microlensing event occurring on two source stars.  A microlensing event with a binary 
lens and a binary source is very rare, and there were only three confirmed cases before the report of 
a new one in this work.  The previously known 2L2S events include MOA-2010-BLG-117 \citep{Bennett2018}, 
OGLE-2016-BLG-1003 \citep{Jung2017}, and KMT-2019-BLG-0797 \citep{Han2021a}, among which MOA-2010-BLG-117L 
is a planetary system and the lenses of the other events are binaries with roughly equal masses.  Then, 
KMT-2018-BLG-1743 is the fourth 2L2S event, and the second case with a lens identified as a planetary 
system.

We checked the feasibility of measuring the microlens parallax $\pie$ and the angular Einstein radius
$\thetae$, which are the two observables needed to determine the physical lens parameters of the mass
$M$ and distance $\dl$ by
\begin{equation}
M={\thetae \over \kappa \pie};\qquad
\dl = {{\rm AU}  \over \pie\thetae+\pi_{\rm S} }.
\label{eq2}
\end{equation}
Here $\kappa=4G/(c^2{\rm AU})$ and $\pi_{\rm S} ={\rm AU}/D_{\rm S}$ represents the parallax of 
the source located at a distance $D_{\rm S}$ \citep{Gould2000}. For the $\pie$ determination, it is 
required to measure the deviation of the lensing light curve from a rectilinear form caused by the 
orbital motion of Earth around the Sun \citep{Gould1992}.  From the additional modeling considering 
the microlens-parallax effect, we found that it was difficult to securely determine $\pie$, because 
the event time scale ($\te\sim 28$~days) was not long enough, and the photometric precision was not 
high enough to firmly detect subtle deviations induced by this second-order effect.  Constraining 
the source orbital motion is even more difficult because the contribution  of the flux from $S_2$ 
is confined in the small region around the peripheral anomaly.  For the estimation of $\thetae$, 
it is required to measure the normalized source radius $\rho$, which is determined from the deviation 
in the caustic-crossing parts of the lensing light curve caused by finite-source effects. With the 
measured $\rho$, the angular Einstein radius was determined by
\begin{equation}
\thetae={\theta_*\over \rho}.
\label{eq3}
\end{equation}
For both the close and wide 2L2S solutions, it was found that the normalized radius of the primary 
source, $\rho_1$, was measured, although the normalized radius of the secondary source, $\rho_2$, 
could not be securely measured.  We note that, although $\rho_2$ is not measured, the angular Einstein 
radius can be determined by $\thetae=\theta_{*,1}/\rho_1$ with the measurement of the angular stellar 
radius of $S_1$, $\theta_{*,1}$. We will describe the detailed procedure of determining the $\theta_{*,1}$ 
and $\thetae$ in the following section.

\begin{table*}[t]
\small
\caption{Source color and magnitude\label{table:four}}
\begin{tabular}{lll}
\hline\hline
\multicolumn{1}{c}{Quantity}   &
\multicolumn{1}{c}{Wide}       &
\multicolumn{1}{c}{Close}      \\
\hline
$(V-I, I)_{\rm RGC}   $     &  (1.861, 15.431)                        &  $\leftarrow$                              \\
$(V-I, I)_{\rm RGC,0} $     &  (1.060, 14.197)                        &  $\leftarrow$                              \\
\hline
$(V-I, I)_{S_1}    $        &  $(1.502 \pm 0.085, 20.829 \pm 0.055)$  &   $(1.479 \pm 0.085, 20.832 \pm 0.057)$    \\
$(V-I, I)_{S_1,0}  $        &  $(0.701 \pm 0.085, 19.595 \pm 0.055)$  &   $(0.678 \pm 0.085, 19.598 \pm 0.057)$    \\
\hline      
$(V-I, I)_{S_2}   $         &  $(3.450 \pm 1.557, 23.544 \pm 0.136)$  &   $(4.893 \pm 5.379, 23.464 \pm 0.139$)    \\
$(V-I, I)_{S_2,0} $         &  $(2.649 \pm 1.557, 22.310 \pm 0.136)$  &   $(4.082 \pm 5.379, 22.230 \pm 0.139$)    \\
\hline      
$q_{F,I}    $               &  $0.082 \pm 0.010$                      &   $0.089 \pm 0.011$                        \\
$q_{F,V}    $               &  $0.014 \pm 0.019$                      &   $0.004 \pm 0.019$                        \\
\hline
\end{tabular}
\tablefoot{ 
The notation ``$\leftarrow$'' indicates that the value is same as in the second column.
}
\end{table*}

\section{Source stars and angular Einstein radius}\label{sec:four}

We estimated the angular Einstein radius using the relation in Eq.~(\ref{eq3}) with the angular radius 
of the source estimated from its color and brightness.  To estimate the reddening and extinction 
corrected (de-reddened) source color and brightness, $(V-I, I)_0$, we used the method of \citet{Yoo2004}, 
in which the centroid of the red giant clump (RGC) in the CMD is used as a reference for calibration.

\begin{figure}[t]
\includegraphics[width=\columnwidth]{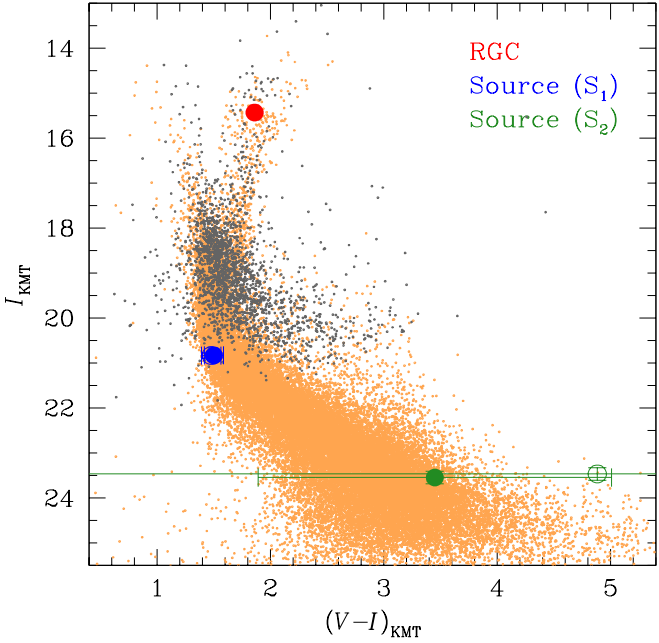}
\caption{
Locations of the primary ($S_1$) and secondary ($S_2$) source stars in the instrumental color-magnitude 
diagram (CMD).  The pair of the filled blue and green dots denote the positions of $S_1$ and $S_2$ 
based on the wide 2L2S solution, and the pair of the empty dots indicate the positions based on the 
close 2L2S solution.  The red filled dot denotes the centroid of red giant clump (RGC).  The {\it Hubble 
Space Telescope} CMD (brown dots) is presented to  show the source locations in the main-sequence branch.
}
\label{fig:twelve}
\end{figure}

Figure~\ref{fig:twelve} shows the locations of $S_1$ and $S_2$ with respect to the RGC centroid (red dot) 
in the instrumental CMD of neighboring stars around the source constructed using the pyDIA photometry 
of the KMTC $I$- and $V$-band data.  The pair of the filled blue and green dots denote the positions of 
$S_1$ and $S_2$ based on the wide 2L2S solution, and the pair of the empty dots indicate the positions 
based on the close 2L2S solution.  It shows that the locations of $S_1$ estimated from the two solutions 
are nearly identical, and the locations of $S_2$ are consistent within the uncertainty.  We also present 
the {\it Hubble Space Telescope} ({\it HST}) CMD \citep[][brown dots]{Holtzman1998} to show the source 
locations in the main-sequence branch. In order to determine the locations of $S_1$ and $S_2$, we first 
estimated the combined flux from the source stars, $F_{S,p}=F_{S_1,p}+F_{S_2,p}$, by conducting a 2L2S 
modeling including the $I$- and $V$-band pyDIA reduction of the data, and then estimated the flux values 
of the individual source stars by
\begin{equation}
F_{S_1,p} = \left( {1\over 1+q_{F,p}}  \right)F_{S,p};\qquad
F_{S_2,p} = \left( {q_{F,p}\over 1+q_{F,p}}  \right)F_{S,p}.
\label{eq4}
\end{equation}
Here the subscript ``$p$'' denotes the passband of the observation, and $q_{F,p}=F_{S_2,p}/F_{S_1,p}$ 
represents the flux ratio between the binary source stars measured in each passband. In Table~\ref{table:four}, 
we list the measured instrumental colors and magnitudes of the RGC centroid, $(V-I, I)_{\rm RGC}$, the 
primary source, $(V-I, I)_{S_1}$, and the secondary source, $(V-I, I)_{S_2}$, together with the flux 
ratios measured in the $I$ and $V$ bands, $q_{F,I}$ and $q_{F,V}$, respectively. We note that $q_{F,I}$ 
values presented in Table~\ref{table:four}, which are derived from the pyDIA reduction, and Table~\ref{table:three}, 
which is derived from the pySIS reduction, are slightly different due to the use of the data from different 
reductions.

With the measured instrumental values, we calibrated the color and brightness of each source using
the offsets in color and magnitude from the RGC centroid, $\Delta (V-I, I)$, by
\begin{equation}
(V-I, I)_0 = (V-I, I)_{{\rm RGC},0} + \Delta (V-I, I), 
\label{eq5}
\end{equation}
where $(V-I, I)_{{\rm RGC},0}=(1.060, 14.197)$ are the de-reddened color and magnitude of the RGC 
centroid known from \citet{Bensby2013} and \citet{Nataf2013}, respectively.  
This relation assumes that the source follows the same distribution of RGC stars.
Under this assumption, the source and RGC centroid experience similar reddening and extinction,
but we note that the granularity of the extinction may cause deviation from this approximation.
We list the de-reddened colors and magnitudes of $S_1$ and $S_2$ in Table~\ref{table:four}.  The 
measured colors and magnitudes indicate that the primary source is an early G-type dwarf, and the 
secondary source is a mid M-type dwarf.

We then estimated the angular radius of the primary source, $\theta_{*,1}$, based on the 
measured de-reddened color and magnitude. We did not estimate the radius of the secondary source 
not only because its normalized radius $\rho_2$ was not measured but also because the uncertainty 
of the measured color was very big.\footnote{Due to the low cadence of $V$-band observations, there 
is only one $V$-band point during the peripheral anomaly (at ${\rm HJD}^\prime=8258.66$) when $S_2$ 
is not significantly brighter than $S_1$. Hence, the constraint on the color of $S_2$ is poor.} The 
angular source radius was estimated from the $(V-K)$--$\theta_*$ relation of \citet{Kervella2004}, where 
$V-K$ color was interpolated from $V-I$ using the color--color relation of \citet{Bessell1988}.\footnote{We 
note that the empirical Kervella relation, expressed by $\log \theta_*=0.5170+0.2755 (V-K)-0.2V$, does 
not require a source distance for the estimation of $\theta_*$.} With the measured source radius, the 
angular Einstein radius and the relative lens-source proper motion were determined by
$\thetae=\theta_{*,1}/\rho_1$ and $\mu =\thetae/\te$, respectively. In Table~\ref{table:five}, we list 
the $(\theta_{*,1}, \thetae, \mu)$ values estimated from the wide and close solutions.  The uncertainties 
of $\theta_*$ and subsequent values of $\thetae$ and $\mu$ are estimated not only based on the uncertainty 
of the measured source color, but also by adding $\sim 7\%$ error in quadrature to account for the scatter 
of RGC stars in the CMD caused by varying distance as well as the uncertainty arising in the 
$(V-K)$--$\theta_*$ process.  We note that the estimated angular Einstein radius of the wide solution, 
$\theta_{\rm E,wide}\sim 0.18$~mas, is substantially smaller than that of the close solution, 
$\theta_{\rm E,close}\sim 0.31$~mas, because the measured normalized source radius, $\rho_{\rm wide} 
\sim 2.1\times 10^{-3}$, is bigger than that of the close solution, $\rho_{\rm close}\sim 1.2\times 10^{-3}$.

\begin{table}[t]
\small
\caption{Angular source and Einstein radii and relative lens-source proper motion\label{table:five}}
\begin{tabular*}{\columnwidth}{@{\extracolsep{\fill}}lll}
\hline\hline
\multicolumn{1}{c}{Quantity}   &
\multicolumn{1}{c}{Wide}       &
\multicolumn{1}{c}{Close}      \\
\hline
$\theta_{*,1}$ ($\mu$as))       &  $0.38 \pm 0.04$   &  $0.36 \pm 0.04$   \\
$\thetae$       (mas)           &  $0.18 \pm 0.04$   &  $0.31 \pm 0.12$   \\
$\mu$ (mas~yr$^{-1}$)           &  $2.34 \pm 0.50$   &  $4.06 \pm 1.61$   \\
\hline
\end{tabular*}
\end{table}

\section{Physical lens parameters}\label{sec:five}

Not being able to measure the microlens parallax, we estimated the physical lens parameters by conducting 
a Bayesian analysis with the use of the constraints provided by the measured observables of the event 
time scale and the angular Einstein radius.

In the Bayesian analysis, we first conducted a Monte Carlo simulation using a prior Galactic model to
generate a large number ($2\times 10^7$) of lensing events. The Galactic model defines the mass function,
physical distribution, and motion of Galactic objects. We used the Galactic model constructed by
\citet{Jung2021}. In summary, disk and bulge objects according to this Galactic model are distributed 
following \citet{Robin2003} and \citet{Han2003} models, respectively, and they move following the bulge 
dynamical model of \citet{Jung2021} and the disk dynamical model of \citet{Han1995}. The masses follow 
the mass function of \citet{Jung2018} commonly for the disk and bulge objects.  In the Bayesian analysis, 
we assume that the physical distribution of stars does not depend on the brightness, that is, faint and 
bright stars are located according to a common distribution.  With the events produced by the simulation, 
the posterior distributions of $M$ and $\dl$ are obtained by constructing their distributions for events 
with $\te$ and $\thetae$ values located within their ranges of uncertainty.

\begin{figure}[t]
\includegraphics[width=\columnwidth]{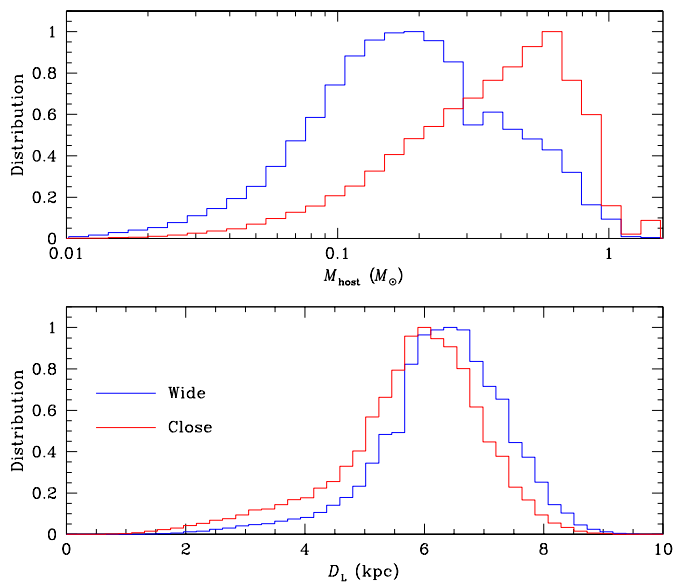}
\caption{
Posterior distributions of the host mass (upper panel) and
distance (lower panel) to the lens constructed from the Bayesian
analysis. The blue and red curves are the distributions
estimated based on the wide and close solutions,
respectively.
}
\label{fig:thirteen}
\end{figure}

Figure~\ref{fig:thirteen} shows the posterior distributions of the host mass $M_{\rm host}\equiv M_1$
(upper panel) and distance (lower panel)
to the planetary lens system. In Table~\ref{table:six}, we list the estimated values 
of $M_{\rm host}$, $M_{\rm planet}\equiv M_2$, $\dl$, and $a_\perp$, where $a_\perp=s \dl \thetae$ 
is the physical projected separation between the planet and host. For each lens parameter, the 
median value is presented as a representative value, and the range of the uncertainty is estimated 
as the 16\% and 84\% of the distribution. Due to the difference in $\thetae$, the lens masses 
estimated from the wide and close solutions are substantially different from each other: 
$(M_{\rm host}, M_{\rm planet}) \sim (0.19~M_\odot, 0.25~M_{\rm J})$ for the wide solution and 
$\sim (0.42~M_\odot, 1.6~M_{\rm J})$ for the close solution. In contrast, the distances estimated 
by the two solutions are similar to each other.

\begin{table}[t]
\small
\caption{Physical lens parameters\label{table:six}}
\begin{tabular*}{\columnwidth}{@{\extracolsep{\fill}}lll}
\hline\hline
\multicolumn{1}{c}{Quantity}   &
\multicolumn{1}{c}{Wide}       &
\multicolumn{1}{c}{Close}      \\
\hline
$M_{\rm host}$ ($M_\odot$)        &  $0.193^{+0.268}_{-0.107}$   &  $0.418^{+0.336}_{-0.251}$   \\
$M_{\rm planet}$ ($M_{\rm J}$)    &  $0.245^{+0.339}_{-0.135}$   &  $1.612^{+1.295}_{-0.967}$   \\
$\dl$ (kpc)                       &  $6.483^{+0.936}_{-1.026}$   &  $6.039^{+0.930}_{-1.272}$   \\
$a_\perp$ (AU)                    &  $1.449^{+0.209}_{-0.229}$   &  $1.931^{+0.297}_{-0.407}$   \\
\hline
\end{tabular*}
\end{table}

\section{Resolving the degeneracy}\label{sec:six}

The degeneracy between the wide and close solutions could have been broken if the peak of 
the light curve had been covered from followup observations.  Due to the high chance of 
planetary perturbations near the peaks of light curves, high-magnification events were 
important targets for intensive follow-up observations in the microlensing experiments 
conducted in a survey+followup mode, in which a survey experiment with a low observational 
cadence focused on finding events and followup teams, for example, RoboNet \citep{Tsapras2009}, 
MiNDSTEp \citep{Dominik2010}, $\mu$FUN \citep{Gould2006}, and ROME/REA \citep{Tsapras2019}, 
conducted intensive observations for the alerted events.  In this mode of lensing experiments, 
it was necessary to monitor lensing events found from the survey to select target events for 
intensive followup observations. This required prodigious human efforts because it was difficult 
to determine which events would have high magnifications based on the sparse data from the surveys.  
With the advent of the high-cadence KMTNet survey using globally distributed telescopes, it is 
now possible to identify high-magnification events with much less efforts for monitoring, enabling 
followup observations that can increase the number of planets found in high-magnification events, 
for example, the Earth-mass planet KMT-2020-BLG0414Lb detected from the combined observations by 
the KMTNet+MOA surveys and LCO \& $\mu$FUN Follow-Up Team \citep{Zang2021}.  Unfortunately, the 
event KMT-2018-BLG-1743 occurred before the operation of the KMTNet AlertFinder system 
\citep{Kim2018}, which became fully operational since the 2019 season, and thus a high-magnification 
alert could not be issued.

Although difficult with the current photometric data, it will be possible to break the degeneracy
by resolving the lens and source from future high-resolution imaging observations using {\it HST} 
or AO system mounted on large ground-based telescopes \citep{Bennett2007}. This is possible because 
the relative lens-source proper motions expected from the wide, $\mu_{\rm wide}\sim 2.3$~mas~yr$^{-1}$, 
and the close, $\mu_{\rm close}\sim 4.1$~mas~yr$^{-1}$, solutions are considerably different from each 
other. For the case of the planetary lensing event OGLE-2005-BLG-169, the lens and source were resolved 
from the {\it HST} \citep{Bennett2015} and the Keck AO imaging \citep{Batista2015} observations, when the 
lens was separated from the source by $\sim 49$~mas.  Using the same criterion, the lens and source of 
KMT-2018-BLG-1743 will be separated about 12 years after the event according to the close solution, that 
is, in 2030.  Unless the lens and source are resolved by that time, the wide solution is more likely to 
be the correct interpretation of the planetary system.  Assuming that the extinction to the lens is 
approximated as $(\dl/\ds)A_I$, where $A_I\sim 0.84$ is the extinction to the source, the expected 
$I$-band brightness of the lens is $I\sim 23.9$ and $\sim 22.8$ according to the wide solution and 
the close solution, respectively.

\section{Summary and conclusion}\label{sec:seven}

We analyzed the microlensing event KMT-2018-BLG-1743 as part of a project in which anomalous 
events with no suggested solutions were reinvestigated among the previous lensing events detected 
in and before the 2019 season by the KMTNet survey.  It was found that the light curve of 
KMT-2018-BLG-1743, which was characterized by a very high peak magnification of $A_{\rm peak}\sim 800$ 
and dual anomalies around the peak and the falling side of the light curve, could not be precisely 
explained by a 2L1S interpretation.  In order to explain the anomalies, we conducted additional 
modeling with the addition of an extra lens and an extra source to a 2L1S interpretation.  From 
this investigation, we found that 2L2S interpretations with a planetary lens system and a binary 
source best explained the observed light curve with $\Delta\chi^2\sim 188$ and $\sim 91$ over the 
2L1S and 3L1S solutions, respectively, giving the event the titles of the fourth 2L2S event and 
the second 2L2S planetary event.  The 2L2S interpretations were subject to  a degeneracy,
mostly caused by the incomplete coverage of the peak region, resulting in two solutions with $s>1.0$ 
and $s<1.0$.  It was found that the source was a binary composed of an early G dwarf and a mid M dwarf.  
The masses of the lens components and the distance to the lens estimated from a Bayesian analysis were 
$(M_{\rm host}, M_{\rm planet}, D_{\rm L})\sim (0.19~M_\odot, 0.25~M_{\rm J}, 6.5~{\rm kpc})$ and 
$\sim (0.42~M_\odot, 1.61~M_{\rm J}, 6.0~{\rm kpc})$ according to the wide and close solutions, 
respectively.  We predicted that the degeneracy between the two solutions would be lifted by 
resolving the lens and source from future high-resolution imaging observations, due to the 
considerable difference in the values of the relative lens-source proper motion expected from 
the two degenerate solutions.

\begin{acknowledgements}
Work by C.H. was supported by the grants of National Research Foundation of Korea (2019R1A2C2085965 
and 2020R1A4A2002885).  This research has made use of the KMTNet system operated by the Korea 
Astronomy and Space Science Institute (KASI) and the data were obtained at three host sites of 
CTIO in Chile, SAAO in South Africa, and SSO in
Australia.

\end{acknowledgements}

\end{document}